\documentstyle[]{article}
\title{ Travelling waves in multivortex configurations }
\author{Jacek Dziarmaga  \\
        Jagellonian University, Institute of Physics,  \\
        Reymonta 4,30-059 Krak\'ow, Poland
        \thanks{Permanent address. E-mail: ufjacekd@ztc386a.if.uj.edu.pl}\\
                        and                            \\
        Institute for Theoretical Physics, Utrecht University,
        \thanks{E-mail: jacek@fys.ruu.nl}                  \\}
\date{20 March 1994}
\begin{document}
\maketitle

    \begin{abstract}

We investigate zero modes of a multivortex background which are due to the
energetic degeneracy of the planar static n-vortex solution in
Bogomol'nyi limit of the Abelian Higgs model parametrised by a set
of 2n real parameters. The zero modes take the form of waves travelling
with a speed of light independently along each of the vortices irrespectively
to their mutual separations. String description of these zero modes is
constructed. We also comment on the remnants of the modes a little outside
of the Bogomol'nyi limit.

    \end{abstract}

\section*{Introduction}

  One of the properties of the soliton solutions is that they break Poincar\'e
symmetries of a vacuum. For example the vortex solution in $2+1$ dimensional
Abelian Higgs model \cite{nielsen} breaks translational invariance. When
extended
to $3+1$ dimensions it also breaks part of spacial rotations. Whenever
the action of the broken symmetry creates out of the given solution
a new solution with different free parameters but of the same energy we can
expect zero modes associated with this symmetry. In
\cite{vachaspati1,vachaspati2} travelling
waves on straight linear vortex background were constructed which are due
to the broken transversal translations. Also broken rotational symmetries
are the origin of a special kind of massless excitations of the vortex
background \cite{khariton}.

  In the Abelian Higgs model there is such a special choice of the coupling
constants - Bogomol'nyi limit \cite{bogomolny,weinberg} at which there are
multivortex
static planar solutions parametrised by a set of $2n$ real parameters
\cite{taubes,weinberg,samols}. These parameters can be choosen for example
as Cartesian coordinates
of the zeros of the Higgs field. It is usually said that these n vortices
do not interact. In fact there are no net static forces between them but
they do interact by forces which are nonzero for nonzero vortex velocities
and are known to lead to nontrivial scattering patterns \cite{ruback,samols,
rubsam,rebbi,ja}. The reasoning
of Vachaspatis \cite{vachaspati1} and also construction
in \cite{vachaspati2} when applied to the
static multivortex solution leads only to overall translational modes
in the form of travelling waves - it makes use of only 2 out of 2n free
parameters. There seems to be a wider possibility to excite independently
the whole set of 2n parameters. Such an excitation will in general
change the intervortex distances. The goal of this paper is to show that it
is indeed possible irrespecive of the fact that there are nontrivial
velocity-dependent interactions between vortices. We also add a comment
on what are the remnants of these excitations outside of the Bogomol'nyi
limit and construct string description of both scattering of parallel
vortices and splitting modes of vortex with winding number 2.

\section{General considerations}

   The Lagrangian of the Abelian Higgs model reads
\begin{equation}\label{gen.10}
  L=-\frac{1}{4}F_{\mu\nu}F^{\mu\nu}+\frac{1}{2}D_{\mu}\phi^{\star}D^{\mu}\phi
                             -\frac{1}{8}\lambda(\phi^{\star}\phi-1)^{2} \;\;,
\end{equation}
where $D_{\mu}\phi=\partial_{\mu}\phi-iA_{\mu}\phi$. The corresponding
field equations are
\begin{equation}\label{gen.20}
  D_{\mu}D^{\mu}\phi=-\frac{1}{2}\lambda(\phi^{\star}\phi-1)\phi \;\;,
\end{equation}
\begin{equation}\label{gen.30}
  \partial_{\mu}F^{\nu\mu}=j^{\nu}\equiv
                  \frac{1}{2}i(\phi^{\star}D^{\nu}\phi-\;c.c.\;) \;\;.
\end{equation}
It is known that at the critical value of the coupling constant of
$\lambda=1$ static version of these equations can be reduced to first order
Bogomol'nyi equations. They are known to possess static multivortex
solutions, which for the topological index n are parametrised by a set
of 2n real parameters (we take n positive for definiteness)
\begin{equation}\label{gen.40}
 \phi=\phi(x,y;\xi_{A}] \;\;\;,\;\;\; A_{\beta}=A_{\beta}(x,y;\xi_{A}] \;\;\;,
                                                          \;\;\; A_{k}=0 \;\;,
\end{equation}
where $\xi_{A}$-s are the parameters. By $\alpha,\beta$... we denote
indices 1,2 while by $i,k$... indices 0,3. The energy of the solution does not
depend on the choosen values of parameters $\xi_{A}$. Any such given
configuration possesses 2n zero modes \cite{weinberg} due to the variation of
2n parameters
\begin{equation}\label{gen.50}
 \delta\phi_{A}=\frac{\partial}{\partial\xi_{A}}\phi(x^{\gamma};\xi_{B}]
                            \;\;\;,\;\;\;
 \delta A_{\beta}^{(A)}=\frac{\partial}{\partial\xi_{A}}
                                      A_{\beta}(x^{\gamma};\xi_{B}] \;\;.
\end{equation}
They were explicitely calculated for the axially symmetric vortex with winding
number n \cite{weinberg}. We would like to show that when we extend
the solution (\ref{gen.40})
to $3+1$ dimensions there are $2n$ kinds of excitations travelling along
the z-axis with the speed of light. We start by very general considerations
without specifying explicitely the set of parameters $\xi_{A}$ at first
and gradually we will become more specific.

  Let us assume that the only change in the Higgs field and in x,y components
of the gauge potential is due to the $t,z$ dependence of the parameters
\begin{equation}\label{gen.60}
  \phi=\phi(x^{\alpha};\xi_{A}(t,z)] \;\;\;,\;\;\;
                                  A_{\beta}(x^{\alpha};\xi_{A}(t,z)] \;\;.
\end{equation}
It is no longer a solution of the field equations (\ref{gen.20},\ref{gen.30}).
To make it a solution we will have to introduce non-zero time dependent
$A_{k}$ components of the gauge potential.

   The first of the field equations (\ref{gen.20}) can be rewritten as
\begin{equation}\label{gen.70}
  D_{i}D^{i}\phi=D_{\beta}D_{\beta}\phi
                            -\frac{1}{2}(\phi^{\star}\phi-1)\phi\;\;.
\end{equation}
The RHS of the above equation is obviously satisfied by the modified
fields (\ref{gen.60}) at any instant of time and at any value of z. The
LHS can be rewritten as
\begin{equation}\label{gen.80}
 \sum_{A}\frac{\partial\phi}{\partial\xi^{A}}\partial_{i}\partial^{i}\xi^{A}+
 \sum_{A,B}\frac{\partial^{2}\phi}{\partial\xi^{A}\partial\xi^{B}}
                           \partial_{i}\xi^{A}\partial^{i}\xi^{B}=
 i(\partial_{i}A^{i})\phi
 +2iA^{i}\sum_{A}\frac{\partial\phi}{\partial\xi^{A}}\partial_{i}\xi^{A}
 +A_{i}A^{i}\phi                                                 \;\;.
\end{equation}
Since we are looking for zero modes we impose on $\xi^{A}$ following two
equations
\begin{equation}\label{gen.90}
 \partial_{i}\partial^{i}\xi^{A}=0 \;\;\;,\;\;\;
 \partial_{i}\xi^{A}\partial^{i}\xi^{B}=0
\end{equation}
for any value of $A,B$. They mean that each $\xi^{A}$ is a function of
$(t+z)$ or $(t-z)$ only. The situation that part of the parameters are
left-movers and another part are right-movers is excluded.
\begin{equation}\label{gen.100}
 \xi^{A}=\xi^{A}(t-z) \;\;\;or\;\;\; \xi^{A}=\xi^{A}(t+z)
\end{equation}
for any A=1,..,2n. Thus we are left only with the RHS
of Eq.(\ref{gen.80}). It can be put identically equal to zero by an Anzatz
\begin{equation}\label{gen.110}
  A_{k}=\sum_{A} F_{A}(x^{\alpha};\xi_{B}]\partial_{k}\xi_{A}  \;\;.
\end{equation}
which implies the gauge condition: $\partial_{\mu}A^{\mu}=0$ since we have
already $\partial_{\beta}A_{\beta}=0$.
Only the coordinate dependence of the functions $F_{A}$ is specified so far.
Otherwise they are arbitrary. With this form of the gauge potential
the field strenght components take the form
\begin{eqnarray}\label{gen.120}
F^{ik}&=&0  \;\;, \nonumber \\
F^{\beta k}&=&\sum_{A}(\partial^{k}\xi^{A})
(\frac{\partial F_{A}}{\partial x_{\beta}}
            -\frac{\partial A^{\beta}}{\partial \xi^{A}}) \;\;,\nonumber \\
F^{\beta\alpha}&=&\partial^{\beta}A^{\alpha}-\partial^{\alpha}A^{\beta} \;\;.
\end{eqnarray}
With the second of these formulas one can easily check that Eq.(\ref{gen.30})
is identically fulfilled for $\nu=\beta$ and it does not impose any extra
constraint on $F_{A}$. It is not so trivial with the case of $\nu=i$. Now
Eq.(\ref{gen.30}) can be rewritten as
\begin{equation}\label{gen.130}
 \sum_{A}(\partial^{i}\xi_{A})
                    (\partial_{\beta}\partial_{\beta}F_{A})=
 \sum_{A}(\partial^{i}\xi^{A})(\phi^{\star}\phi)
                   [F_{A}-\frac{\partial{\Theta}}{\partial\xi^{A}}]      \;\;.
\end{equation}
Since the dependence of particular $\xi^{A}$ on $x^{i}$ can be choosen
arbitrarily this equation can be further reduced to
\begin{equation}\label{gen.140}
  (\frac{\partial}{\partial x^{\beta}}\frac{\partial}{\partial x^{\beta}}
   -\phi^{\star}\phi)\; F_{A}=
   -(\phi^{\star}\phi)\frac{\partial\Theta}{\partial\xi^{A}}  \;\;,
\end{equation}
where $\Theta$ is an actual value of the phase of the Higgs field,
$\phi=\mid\phi\mid\exp i\Theta$. It is a static equation for any given
value of the parameters $\xi^{A}$.

  There is one special case when its solution can be given without any effort
for any value of the parameters. We can decompose the set of parameters
$\xi^{A}$ into a set of two overall translational modes
\cite{vachaspati1}
$X,Y$ and the rest of the modes $\xi^{A},A=1,..,2n-2$, which we will call
"splitting modes". Equation (\ref{gen.60}) now reads
\begin{equation}\label{gen.150}
  \phi=\phi(x+X,y+Y;\xi^{A}] \;\;\;,\;\;\;
  A_{\beta}=A_{\beta}(x+X,y+Y;\xi^{A}] \;\;,
\end{equation}
Equation (\ref{gen.140}) for the translational modes takes the form
\begin{equation}\label{gen.160}
  (\partial_{\beta}\partial_{\beta}-\phi^{\star}\phi)F_{(x^{\beta})}=
  -\phi^{\star}\phi\frac{\partial\Theta}{\partial x^{\beta}}
\end{equation}
and its exact solution is
\begin{equation}\label{gen.170}
F_{(x^{\beta})}=A_{\beta}(x^{\alpha}+X^{\alpha};\xi_{A}] \;\;.
\end{equation}
This particular solution is known thanks to Vachaspatis \cite{vachaspati1},
they are
overall translational modes. The general form of the gauge potential now reads
\begin{equation}\label{gen.180}
A_{k}=\sum_{\beta}A_{\beta}(x^{\alpha}+X^{\alpha};\xi^{B}]
                                      \partial_{k}X^{\beta}+
\sum_{A=1}^{2n-2}F_{A}(x^{\alpha}+X^{\alpha};\xi_{B}]\partial_{k}\xi_{A}\;\;.
\end{equation}
When we shift $x^{\alpha}$ coordinates
$x^{\alpha}\rightarrow y^{\alpha}=x^{\alpha}+X^{\alpha}$,
we can write Eq.(\ref{gen.140}) for the splitting modes as
\begin{equation}\label{gen.190}
(\frac{\partial}{\partial y^{\beta}}\frac{\partial}{\partial y^{\beta}}
 -\phi^{\star}\phi)\; F_{A}(y^{\alpha};\xi^{B}]=
 -(\phi^{\star}\phi)\frac{\partial \Theta}{\partial\xi^{A}} \;\;,
\end{equation}
which is formally the same as the original equation (\ref{gen.140}).
Thus we can see that the splitting modes can be investigated on the background
of the translational modes. The only effect of the Vachaspatis modes is the
time dependent shift of the coordinates in Eq.(\ref{gen.190}). Another effect
is a correlation, which states that both translational modes and splitting
modes have to be exclusively right-movers or left-movers. For the analysis
of Eq.(\ref{gen.190}) itself we can forget about translational modes and think
of them as if they were put equal to zero.

\section{ Solutions to the effective equation }

   Analysis of equation (\ref{gen.140}) or (\ref{gen.190}) is quite nontrivial
due to a very limited knowledge about the background field. Neverthelles it
can be completed in certain limit cases.

  In general the n-vortex configuration in $2+1$ dimensions can be
described by \cite{samols}
\begin{equation}\label{sym.10}
  \phi=(z^{n}-\sum_{k=0}^{n-1}\lambda_{k}z^{k})W(z,z^{\star}) \;\;,
\end{equation}
where $z=x+iy=re^{i\theta}$ and $W(z,z^{\star})$ is a positively
definite real function. The $\lambda$-polynomial itself determines both the
phase of the Higgs field and the positions of its zeros. There is one to
one correspondence between the field configuration and the set of parameters
$\lambda_{k}$ \cite{samols}. The phase of the Higgs field is given by:
\begin{equation}\label{sym.20}
  \Theta(z;\lambda_{k}]=Arg(z^{n}-\sum_{k=0}^{n-1}\lambda_{k}z^{k})
                                                                \;\;\;,\;\;\;
  \lambda_{k}=\lambda_{k}^{1}+i\lambda_{k}^{2}  \;\;.
\end{equation}
First we will analyse the limit case of all $\lambda^{\alpha}_{k}=0$.

\subsection{ Case of n coincident vortices }

  In the limit of vanishing $\lambda$-s, which corresponds to
n vortices sitting on top of each other, the desired derivatives of $\Theta$
with respect to the parameters are
\begin{equation}\label{sym.30}
  \frac{\partial\Theta}{\partial\lambda^{1}_{k}}=
  \frac{\sin(m\theta)}{r^{m}} \;\;,\;\;
  \frac{\partial\Theta}{\partial\lambda^{2}_{k}}=
  -\frac{\cos(m\theta)}{r^{m}}\;\;,\;\;m=n-k\;\;,
\end{equation}
When the background field configuration is
$ \phi=f(r)\exp in\theta $, eq.(\ref{gen.140}) can be written in the form
\begin{eqnarray}\label{sym.35}
(\nabla^{2}-f^{2}(r))F_{1k}(r,\theta) & = &
                      -\frac{f^{2}(r)}{r^{m}}\sin(m\theta)  \;\;, \nonumber \\
(\nabla^{2}-f^{2}(r))F_{2k}(r,\theta) & = &
                       \frac{f^{2}(r)}{r^{m}}\cos(m\theta)  \;\;,
\end{eqnarray}
With a decomposition
\begin{equation}
F_{1k}(r,\theta)=F_{k}(r)\sin(m\theta)\;\;\;,\;\;\;
F_{2k}(r,\theta)=-F_{k}(r)\cos(m\theta)\;\;,
\end{equation}
the above equations will be reduced to a single radial equation
\begin{equation}\label{sym.40}
  F_{k}^{\prime\prime}+\frac{1}{r}F_{k}^{\prime}-\frac{m^{2}}{r^{m}}F_{k}
  -f^{2}F_{k}=-\frac{f^{2}}{r^{m}} \;\;,\;\;m=n-k \;\;.
\end{equation}
With a help of an asymptotics $f(r)\sim f_{0}r^{n}$ one can analyse this
equation near the origin $(r\sim 0)$:
\begin{equation}\label{sym.50}
  F_{k}(r)=\frac{a}{r^{m}}+br^{m}+cr^{2m}
           -\frac{f_{0}^{2}}{4(k+1)(n+1)}r^{2k+2}+...  \;\;\;.
\end{equation}
The point is that there is only one singular term $O(\frac{1}{r^{m}})$.
We will show that the coefficient $a$ of this term can be tuned to zero.

    We can regard equation (\ref{sym.40}) as an inhomogenous linear
differential equation. For large $r$ where $f^{2}\sim 1$, the homogenous
part of the equation approaches the modified Bessel equation. We choose the
asymptotically vanishing solution as
\begin{equation}\label{sym.60}
  F^{hom}_{k}(r)\sim d\frac{e^{-r}}{\sqrt{r}} \;\;,\;\;r\rightarrow\infty
                                                                          \;\;.
\end{equation}
Since for small $r$ $f^{2}\rightarrow 0$ and $\frac{m^{2}}{r^{2}}$ diverges,
this asymptotically vanishing solution is transferred into a linear
combination
\begin{equation}\label{sym.70}
  F^{hom}_{k}\sim d(\frac{e}{r^{m}}+...)+d(gr^{m}+...) \;\;,
  \;\;r\rightarrow 0  \;\;.
\end{equation}
$d$ is an overall constant which will be choosen later. We expect
$e$ to be nonzero. An indirect proof of this fact can be extracted
from considerations in \cite{burzlaf}. We need this singular term
to remove an eventual singularity of the total solution.

   On the other hand we can look for a special solution of the whole
inhomogenous equation, which for large $r$ can be solved by
\begin{equation}\label{sym.80}
  F^{inhom}_{k}(r)\sim\frac{1}{r^{m}} \;\;,\;\; r\rightarrow\infty \;\;.
\end{equation}
For small r it is in general transferred into a singularity
\begin{equation}\label{sym.90}
  F^{inhom}_{k}(r)\sim \frac{s}{r^{m}}+...\;\;,\;\; r\rightarrow 0 \;\;.
\end{equation}
If the coefficient $s$ happens to be equal to $0$, we will also choose
$d=0$ and there will be no singularity in Eq.(\ref{sym.50}). If $s\neq 0$,
we will have to choose $de=-s$ and once again singularity will be removed.
Eq.(\ref{sym.50}) shows that there is no other singular term to be removed.
Thus potentials $A_{i}$
\begin{equation}\label{sym.100}
  A_{i}=\sum_{k=0}^{n-1} F_{k}(r)[(\partial_{i}\lambda^{1}_{k})\sin(n-k)\theta
                     -(\partial_{i}\lambda_{k}^{2})\cos(n-k)\theta] \;\;,
\end{equation}
are regular for small r. The asymptotics for large $r$ is dominated
by a contribution from the inhomogenous part: $F_{k}\sim\frac{1}{r^{m}}$
without any undetermined coefficients
\begin{equation}\label{sym.110}
  A_{i}\sim\sum_{m=1}^{n}
       [(\partial_{i}\lambda_{k}^{1})\frac{\sin m\theta}{r^{m}}-
        (\partial_{i}\lambda_{k}^{2})\frac{\cos m\theta}{r^{m}}] \;\;\;,
  \;\;\; m=n-k  \;\;,
\end{equation}
This asymptotic form is an exact result. We expect it to be also valid
for finite values of $\lambda$-s. An asymptotic form of the derivatives
of the phase of the Higgs field for any finite values of $\lambda$-s
is still given by Eq.(\ref{sym.30}). For large $r$ also general
$\phi^{\star}\phi\sim 1$ and the "homogenous" solutions must be taken
to be asymptotically exponentially vanishing. Thus Eq.(\ref{sym.110})
is a general asymptotic result.

  The existence of solutions to equation (\ref{gen.140}) in the limit of all
$\lambda^{\alpha}_{k}=0$ strongly suggests that we can expect them to exist
also for small but finite $\lambda$-s.

\subsection{ Small fluctuations around n-vortex background }

  With a help of the Bogomol'nyi equations at the critical value of the
coupling constant
\begin{equation}\label{sm.10}
  (D_{1}+iD_{2})\phi=0 \;\;\;,\;\;\;F_{12}=\frac{1}{2}(1-\phi^{\star}\phi)
\end{equation}
one can obtain a general perturbation of the Higgs field \cite{ruback,ja}
around the background of n coincident vortices $f(r)e^{in\theta}$ :
\begin{equation}\label{sm.20}
  \phi+\delta\phi=f(r)e^{in\theta}
                  [1+\sum_{s=0}^{n-1}\lambda_{s}H_{n-s}(r)e^{-i(n-s)\theta}]
                  +O(\lambda^{2})         \;\;,
\end{equation}
where functions $H_{p}(r)$ satisfy
\begin{equation}\label{sm.30}
  H^{\prime\prime}_{p}+\frac{1}{r}H^{\prime}_{p}-\frac{p^{2}}{r^{2}}H_{p}-
  f^{2}H_{p}=0 \;\;,
\end{equation}
which happens to be identical with the homogenous part of Eq.(\ref{sym.40}).
We choose the normalisation of the solution (\ref{sym.60},\ref{sym.70})
in such a way that
\begin{equation}\label{sm.40}
  H_{p}(r)\sim -\frac{1}{r^{p}} \;\;\;,\;\;\;r\rightarrow 0 \;,
\end{equation}
With this normalisation $\lambda_{s}$ in Eq.(\ref{sm.20}) can be
identified with those in Eq.(\ref{sym.10},\ref{sym.20}),
because for small $r$ the Higgs field is well approximated by
\begin{equation}\label{sm.45}
  \phi+\delta\phi \approx (z^{n}-\sum_{s=0}^{n-1} \lambda_{s}z^{s}) \;\;
\end{equation}
With Eq.(\ref{sm.20}) the approximate phase of the Higgs field reads
\begin{equation}\label{sm.46}
 \Theta=n\theta+\frac{1}{2i}\ln[\frac
        {1+\sum_{s=0}^{n-1}\lambda_{s}H_{n-s}(r)e^{-i(n-s)\theta}}
        {1+\sum_{l=0}^{n-1}\lambda_{l}^{\star}H_{n-l}(r)e^{i(n-l)\theta}}]
                                                                       \;\;,
\end{equation}
Eq. (\ref{gen.140}) for any particular $\lambda^{\alpha}_{k}$ can be always
transformed into equation for $\lambda^{1}_{k}$ by an appropriate rotation
of the reference frame, so without loss of generality we can restrict to this
case. The derivative of the phase of the Higgs field with respect to
$\lambda^{1}_{k}$ is
\begin{equation}\label{sm.47}
  \frac{\partial\Theta}{\partial\lambda^{1}_{k}}=\frac
  {H_{n-k}(r) \;\Im\; [e^{i(k-n)\theta}
   +\sum_{l=0}^{n-1}\lambda^{\star}_{l}H_{n-l}(r)e^{i(k-l)\theta}]}
  {\mid 1+\sum_{s=0}^{n-1}\lambda_{s}H_{n-s}(r)e^{-i(n-s)\theta} \mid^{2}}
                                                                       \;\;.
\end{equation}
Eq.(\ref{gen.140}) when we make an expansion
$F_{\alpha k}\rightarrow F_{\alpha k}+\delta F_{\alpha k}+O(\lambda^{2})$,
($F_{\alpha k}$ corresponds to $\lambda^{\alpha}_{k}$), takes the form
\begin{equation}\label{sm.70}
  (\nabla^{2}-f^{2}) F_{\alpha k}(r,\theta) =
  -f^{2}\frac{\partial\Theta}{\partial\lambda^{\alpha}_{k}} \;\;,
\end{equation}
\begin{equation}\label{sm.80}
  (\nabla^{2}-f^{2}) \delta F_{\alpha k}(r,\theta) =
  (\delta\phi^{\star}\phi) F_{\alpha k}
  -\delta(\phi^{\star}\phi\frac{\partial\Theta}{\partial\lambda^{\alpha}_{k}})
                                                                         \;\;,
\end{equation}
where the first equation comes from the terms $O(\lambda^{0})$, the second
equation from those linear in $\lambda$-s. We have omitted higher order
equations in this hierachy. The first equation has been already analysed
in section 2.1
\begin{equation}\label{sm.90}
  F_{1k}(r,\theta)=F_{k}(r)\sin(n-k)\theta \;\;\;,\;\;\;
  F_{2k}(r,\theta)=-F_{k}(r)\cos(n-k)\theta \;\;.
\end{equation}
Upon substitution of this solution, Eq.(\ref{sm.80}) once again becomes
linear inhomogenous equation. In this way any equation in the hierarchy,
which begins with (\ref{sm.70},\ref{sm.80}), can be step by step cast in
a form of linear inhomogenous equation of d'Alembert type and when
solved provide a complete input for the next equation in order of the powers
of $\lambda$. This property in principle provides us with a systematic
calculational scheme. Neverthelles for practical calculations for any
given set of parameters we think Eq.(\ref{gen.140}) to be much easier
menegable by numerical methods. We will use the hierarchy only to analyse
solutions for small $\lambda$-s.

  When we take into account that
\begin{equation}\label{sm.100}
 \mid \phi+\delta\phi \mid^{2} =
 f^{2}(r)
 \mid 1+\sum_{s=0}^{n-1}\lambda_{s}H_{n-s}(r)e^{-i(n-s)\theta} \mid^{2}
\end{equation}
Eq.(\ref{sm.80}) in our case can be explicitely written as
\begin{equation}\label{sm.110}
\begin{array}{rl}
(\nabla^{2}-f^{2})\delta F_{1k}(r,\theta)=&
                      2f^{2}F_{k}\sin(n-k)\theta\sum_{s=0}^{n-1}H_{n-s}(r)
           [\lambda^{1}_{s}\cos(n-s)\theta+\lambda^{2}_{s}\sin(n-s)\theta]  \\
          &+f^{2}H_{n-k}\sum_{s=0}^{n-1} H_{n-s}
           [\lambda^{2}_{s}\cos(k-l)\theta-\lambda^{1}_{s}\sin(n-s)\theta]  \\
    \equiv & \sum_{l=0}^{2n} [W_{l}(r)\cos l\theta + U_{l}(r)\sin l\theta]
\end{array} \;\;,
\end{equation}
where we have rewritten the RHS in a form of its most general Fourier
transform in $\theta$. $W_{l}(r)$ and $U_{l}(r)$ are regular functions
quickly vanishing at infinity. Some of them are identically equal to zero.
With a decomposition
\begin{equation}\label{sm.120}
  \delta F_{1k}(r,\theta)=\sum_{l=0}^{2n}
        [g_{l}(r)\cos l\theta + h_{l}(r)\sin l\theta] \;\;,
\end{equation}
we arrive at a set of linear inhomogenous differential equations
\begin{eqnarray}\label{sm.130}
  g_{l}^{\prime\prime}+\frac{1}{r}g_{l}^{\prime}-\frac{l^{2}}{r^{2}}g_{l}
                                                    -f^{2}g_{l}=W_{l} \;\;, \\
  h_{l}^{\prime\prime}+\frac{1}{r}h_{l}^{\prime}-\frac{l^{2}}{r^{2}}h_{l}
                                                    -f^{2}h_{l}=U_{l} \;\;.
\end{eqnarray}
The regularity of their solutions at $r=0$ can be established by a similar
analysis as in section 2.1.

 A natural question arises whether we
can expect regularity of solutions for any finite values of $\lambda$-s.
In principle we could repeat our method of analysis to any finite order
of the hierarchy beginning with (\ref{sm.70},\ref{sm.80}). An indication
that the result of such an analysis can be expected to be positive
is that we can give an approximate solution to Eq.(\ref{gen.140}) for large
separations of vortices.

\subsection{ Well separated vortices }

  Let us concern a situation in which there is  certain number of vortices
labelled by $s$, with winding numbers $n_{s}$ and positions $\vec{R}_{s}$.
This planar configuration is to be regarded as a background input to
Eq.(\ref{gen.140}) so in reality it can quite as well fit to a situation
when at certain value of $z$ at certain moment of time the cross-section
through the vortex spaghetti is a planar configuration of well separated
vortices.

  Vortices are well separated when distances between them
\begin{equation}\label{w.10}
  \mid \vec{R}_{s}-\vec{R}_{s^{\prime}} \mid \gg 1
\end{equation}
for any values of $s$ and $s^{\prime}$. In our dimensionless Lagrangian
the mass scales are of order $1$, so our condition means that solutions
around separate $\vec{R}_{s}$ approach asymptotics very quickly as compared
to the intervortex distances. In this case the background fields can be
approximated by
\begin{eqnarray}\label{w.20}
  \phi(x,y;\vec{R}_{s}] & \approx & \prod_{s} \phi^{(s)}(x-X_{s},y-Y_{s}) \;\;,
                                                                 \nonumber \\
  A_{\beta}(x,y;\vec{R}_{s}] & \approx &
                              \sum_{s} A_{\beta}^{(s)}(x-X_{s},y-Y_{s}) \;\;,
\end{eqnarray}
where $\phi^{(s)}$ and $A_{\beta}^{(s)}$ are fields of separate vortices
with winding numbers $n_{s}$. This is a good approximation up to terms
vanishing exponentially with intervortex distances. With the same degree
of accuracy an approximate solution for $A_{i}$ can be given
\begin{equation}\label{w.30}
  A_{i} \approx -\sum_{s,\beta}(\partial_{i}R^{\beta}_{s})
                                  A_{\beta}^{(s)}(x-X_{s},y-Y_{s})  \;\;.
\end{equation}
Separate $X_{s}$,$Y_{s}$ vary independently due to general results of
section 1. All of them have to be exclusively left-movers or right-movers.

  Putting together all the results of section 2 we think justified expectation
that Eq.(\ref{gen.140}) possesses solution for any value of the parameters
$\xi^{A}$. Let us give three examples of particularly symetric excitations
travelling with a speed of light along vortex with winding number two
$\phi=(z^{2}-\lambda)W(z,z^{\star})$.

  a) Let us take $\lambda=R^{2}\sin(k_{i}x^{i})$, $k_{i}k^{i}=0$.
Corresponding positions of the zeros of the Higgs field are
\begin{equation}\label{ex.10}
  z_{1,2}=\stackrel{+}{-}R\sqrt{\sin(k_{i}x^{i})}  \;\;\;,\;\;\;
  z_{1,2}=\stackrel{+}{-}iR\sqrt{-\sin(k_{i}x^{i})}  \;\;,
\end{equation}
for the positive and negative $\sin(k_{i}x^{i})$ respectively. This solution
describes travelling splitting wave. At the points where the actual amplitude
is equal to zero (coincident vortices) the orientation of the wave is twisted
by the right angle.

  b) Another example is $\lambda=R^{2}e^{2ik_{i}x^{i}}$ which corresponds
to positions of vortices
\begin{equation}\label{ex.20}
  z_{1}=Re^{i(k_{i}x^{i})} \;\;\;,\;\;\; z_{2}=Re^{i(k_{i}x^{i}+\pi)} \;\;,
\end{equation}
This excitation has a form of double helix moving up or down $z$-axis with
a speed of light.

  c) Our final example is $\lambda=R^{2}e^{2i\Theta(k_{i}x^{i})}$, where
$\Theta(k_{i}x^{i})=arctg(k_{i}x^{i})$ interpolates between $\frac{-\pi}{2}$
for negative argument and $\frac{\pi}{2}$ for positive $k_{i}x^{i}$.
Vortices at $R\exp i\Theta$ and $R\exp i(\Theta+\pi)$ are parallel lines
far from $k_{i}x^{i}\approx 0$ and around this point they make a twist
by an angle of $\pi$. The twist moves with a speed of light.

  In all three cases we have taken into account only splitting modes. There
is a wider possibility to take these splitting modes on a translational
background.

\section{String description and adiabatic approximation}

  Vortices in $2+1$ dimensions can be regarded as point-like particles
while those in $3+1$ dimensions were invented \cite{nielsen} to be
field-theoretical realisation of strings. Scattering of vortices on plane
was succesfully described with a help of Manton's approximation
\cite{ruback,samols}
which amounts to reduction of the full field-theoretical dynamics in
n-vortex sector to mechanics of n point particles on curved surface.
For example dynamics of pair of vortices can be described in their center
of mass frame by an effective nonrelativistic Lagrangian
\begin{equation}\label{str.10}
  L_{eff}=h_{\alpha\beta}(\xi)
          \frac{d\xi^{\alpha}}{dt}
          \frac{d\xi^{\beta}}{dt} \;\;,
\end{equation}
where positions of vortices on the plane are $(\xi^{1},\xi^{2})$ and
$(-\xi^{1},-\xi^{2})$. $h_{\alpha\beta}(\xi)$ is a metric tensor, symmetric
with respect to $\xi\rightarrow -\xi$, which incorporates effects of
vortex mutual interactions. For large separations it is asymptotically flat
$h_{\alpha\beta}\approx \delta_{\alpha\beta}$ while for small separations
it tends to $R^{2}(\dot{R}^{2}+R^{2}\dot{\Theta}^{2})$ in polar coordinates
and is responsible for the right-angle scattering in the head-on collision.
This approximation gives results in accordance with direct numerical
simulations up to velocity 0.4 \cite{samols}. The effective Lagrangian
(\ref{str.10}) was derived in adiabatic approximation so an infinite
series of terms of higher order in velocity was systematically neglected.
We would like now turn attention to the $3+1$ dimensional case and unify
this nonrelativistic theory with our essentially relativistic results
presented in previous sections.

   Translational waves on an isolated vortex coincide with analogous
excitations of the Nambu-Goto string \cite{vachaspati1}. Let us imagine
two strings parametrised for example by
\begin{equation}\label{str.20}
 \xi^{\mu}_{(1)}=[\tau,\xi^{1}(\tau,\sigma),\xi^{2}(\tau,\sigma),\sigma]
                                                             \;\;\;,\;\;\;
 \xi^{\mu}_{(2)}=[\tau,-\xi^{1}(\tau,\sigma),-\xi^{2}(\tau,\sigma),\sigma]
                                                                    \;\;.
\end{equation}
A point on $z$-axis is a "center-of-mass" for each cross section by a plane
of constant $z$ through this configuration of two strings. The action
for the Nambu-Goto string is constructed out of a metrics induced
on its world-sheet, namely
$\gamma_{ik}=\eta_{\mu\nu}\frac{\partial\xi^{\mu}}{\partial\tau^{i}}
                          \frac{\partial\xi^{\nu}}{\partial\tau^{k}}$.
In this way we would obtain two Nambu-Goto strings with nontrivial
mutual interactions. How to geometrise these interactions ? Let us take
a cross section through our system at any fixed $z$, thus we obtain
2-particle mechanical system in $C-M$ frame. We propose to replace Minkowskian
metrics in definition of $\gamma_{ik}$ by an effective metrics
which incorporates mutual interactions
\begin{equation}\label{str.30}
    G_{\mu\nu}=\left[   \begin{array}{cc}
                                  \eta_{ik} &      0                       \\
                                      0     & -h_{\alpha\beta}(x^{\gamma})
                      \end{array}
             \right ]
\end{equation}
The effective induced metrices now read
\begin{equation}\label{str.40}
  g^{(1)}_{ik}=\eta_{mn}\frac{\partial\xi^{m}}{\partial\tau^{i}}
                        \frac{\partial\xi^{n}}{\partial\tau^{k}}
             -h_{\alpha\beta}(\xi)\frac{\partial\xi^{\alpha}}{\partial\tau^{i}}
                                  \frac{\partial\xi^{\beta}}{\partial\tau^{k}}
             =g^{(2)}_{ik} \equiv g_{ik} \;\;,
\end{equation}
The effective two string action becomes
\begin{equation}\label{str.50}
  S_{eff}=-\pi\int d\tau d\sigma \sqrt{-g^{(1)}}
          -\pi\int d\tau d\sigma \sqrt{-g^{(2)}}=
         -2\pi\int d\tau d\sigma \sqrt{-g}         \;\;,
\end{equation}
where $g=det\{ g_{ik} \}$ and prefactors $\pi$ are string tensions equal to
linear energy density of a vortex in Bogomol'nyi limit.

   Let us test this construction on two examples. First one is a case of slow
motion of parallel strings: $\xi^{\alpha}=\xi^{\alpha}(\tau)$. Now the action
(\ref{str.50}) reduces to
\begin{equation}\label{str.60}
  S_{eff}^{2+1}=-2\pi \int d\tau \sqrt{ g_{\tau\tau} } \;\;\;,\;\;\;
   g_{\tau\tau}=(\frac{d\xi^{0}}{d\tau})^{2}
              -h_{\alpha\beta}(\xi) \frac{d\xi^{\alpha}}{d\tau}
                                     \frac{d\xi^{\beta}}{d\tau}  \;\;,
\end{equation}
which is parametrisation invariant. We can introduce a $1$-dim Vielbein
on the worldline
$g_{\tau\tau}=e^{0}_{\tau}\eta_{00}e^{0}_{\tau} \equiv e^{2}$ as an additional
dynamical variable
\begin{equation}\label{str.70}
  S_{eff}^{2+1}=-2\pi\int d\tau [e-\mu(e^{2}-g_{\tau\tau})] \;\;,
\end{equation}
where $\mu$ is a Lagrange multiplier and $g_{\tau\tau}$ is defined in
equation (\ref{str.60}). Variation with respect to $e$ leads to an equation
$\mu=\frac{1}{2e}$. When we fix the gauge $e=1$ and drop constant terms from
the action it will take quadratic form
\begin{equation}\label{str.80}
  S_{eff}^{2+1}=-\pi \int d\tau
                [(\frac{d\xi^{0}}{d\tau})^{2}
                 -h_{\alpha\beta}(\xi)\frac{d\xi^{\alpha}}{d\tau}
                                       \frac{d\xi^{\beta}}{d\tau}]
\end{equation}
with a constraint
\begin{equation}\label{str.90}
  (\frac{d\xi^{0}}{d\tau})^{2}=1+h_{\alpha\beta}(\xi)
                                         \frac{d\xi^{\alpha}}{d\tau}
                                         \frac{d\xi^{\beta}}{d\tau}  \;\;.
\end{equation}
$\xi^{0}$ component is fully determined by the constraint and othewise
decouples, while the dynamics of $\xi^{\alpha}$-s is governed by a Lagrangian
\begin{equation}\label{str.100}
     L_{eff} \simeq h_{\alpha\beta}(\xi)\frac{d\xi^{\alpha}}{d\tau}
                                        \frac{d\xi^{\beta}}{d\tau} \;\;,
\end{equation}
which is identical with (\ref{str.10}) except for parametrisation by
the proper time $\tau$ instead of $t$. From the constraint one can find
relation between $\tau$ and $t$. From (\ref{str.100}) we will obtain
the same geodesic curve as from (\ref{str.10}). The only difference
negligible for small velocities is a different parametrisation. It does not
change deflection angles in scattering of vortices but from the point
of view of the action (\ref{str.60}) there is a different correspondence
between velocity used in adiabatic approximation (\ref{str.100}), which is
velocity with respect to $\tau$ and that in direct numerical simulation
which is velocity with respect to laboratory time $t$. An important difference
arises when we go a little outside of the Bogomol'nyi limit like in
\cite{shah}. In this case due to potential forces between vortices they
no longer follow geodesics but now their scattering angle depends both
on impact parameter and on initial velocity. In the Bogomol'nyi limit up
to velocity 0.4 there is a good agreement between analytic results and
those from numerical simulation \cite{samols} while outside of the critical
coupling there is a systematic discrepancy \cite{shah} vanishing for small
velocities and smoothly rising as they go from 0 to 0.4. This systematic
error can be explained from our point of view. For velocities greater then
0.4 scattering is no longer elastic so point particle approximation in
both cases does not work any longer.

   Thus we can see that this construction fits to $2+1$ dimensional
results but what about travelling waves? Variation of the action
(\ref{str.50}) leads to equation of motion
\begin{equation}\label{str.110}
  \frac{1}{\sqrt{-g}}\frac{d}{d\tau^{i}}
  (\sqrt{-g}g^{mn}\frac{\partial g_{mn}}{\partial\xi^{\alpha}_{,i}})=
  g^{mn}\frac{\partial g_{mn}}{\partial\xi^{\alpha}}   \;\;.
\end{equation}
With a parametrisation (\ref{str.20}) the effective metric reads
\begin{equation}\label{str.120}
  g_{mn}=\eta_{mn}-h_{\alpha\beta}(\xi) \frac{\partial\xi^{\alpha}}
                                             {\partial\tau^{m}}
                                        \frac{\partial\xi^{\beta}}
                                             {\partial\tau^{n}}   \;\;.
\end{equation}
When we impose a condition $\xi^{\alpha}=\xi^{\alpha}(\tau+\sigma)$ or
$\xi^{\alpha}=\xi^{\alpha}(\tau-\sigma)$ exclusively, then one can easily
check that $g=-1$ and thus simplified equation (\ref{str.110}) is identically
fulfilled. This reasoning is not sensitive to the precise form of the
metric $h_{\alpha\beta}$, it is enough to assume for the whole construction
its parity and that it is a symetric invertible matrix. Thus we can see
that also in this construction vortices can be independently translationally
excited and they do not feel interactions essential for the scattering
of parallel strings.

   Now let us make a bold extension of this construction to the case of
slightly curved vortex with winding number 2. Let us define a "center of mass
frame" in this case by
\begin{equation}\label{str.130}
  x^{\mu}=R^{\mu}(\tau,\sigma)+\rho^{\alpha}n^{\mu}_{(\alpha)}(\tau,\sigma)
                                                                       \;\;,
\end{equation}
where $n^{2}_{(\alpha)}=-1$, $\eta_{\mu\nu}n^{\mu}_{(\alpha)}R^{\mu}_{,i}$ and
we assume the reference frame to be torsion-free
$\eta_{\mu\nu}n^{\mu}_{(\alpha)}n^{\nu}_{(\beta),i}=0$. The
strings in coordinates $\{\tau,\sigma,\rho^{\alpha} \}$ are parametrised as
\begin{equation}\label{str.140}
  \xi^{a}_{(1)}(\tau,\sigma)=[\xi^{0},\xi^{1},\xi^{2},\xi^{3}]
                                                                \;\;,\;\;
  \xi^{a}_{(2)}(\tau,\sigma)=[\xi^{0},-\xi^{1},-\xi^{2},\xi^{3}] \;\;,
\end{equation}
what defines the frame used. When $\xi^{\alpha}=0$ both strings coincide
with the worldsheet $R^{\mu}$ or $\rho=0$. For $\xi^{\alpha}\neq 0$ they are
uniformly splitting from $R^{\mu}$. The metric in new coordinates
\begin{equation}\label{str.150}
   G_{ab}= \left[ \begin{array}{cc}
                          \gamma_{ik} &           0            \\
                               0      &  -\delta_{\alpha\beta}
                  \end{array} \right] + O(\rho)
                      \rightarrow
   G_{ab}^{eff}= \left[ \begin{array}{cc}
                          \gamma_{ik} &           0            \\
                               0      &  -h_{\alpha\beta}(\xi)
                  \end{array} \right] + O(\xi)
\end{equation}
is once again replaced by an effective metric incorporating efects of
interactions of separate strings. Terms $O(\xi)$ are small for small external
curvature of the worldsheet $R^{\mu}$ and sufficiently small $\xi^{\alpha}$.
$\gamma_{ik}$ is a metric induced on the line $R^{\mu}(\tau,\sigma)$.
Effective metric tensors are
\begin{equation}\label{str.160}
  g_{ik}^{(1)}=G_{ab}^{eff}(\xi)\frac{\partial\xi^{a}}{\partial\tau^{i}}
                                \frac{\partial\xi^{a}}{\partial\tau^{k}}
              =g_{ik}^{(2)} \equiv g_{ik}
\end{equation}
and they give rise to the effective action
\begin{equation}\label{str.170}
  S_{eff}=-2\pi\int d\tau d\sigma \sqrt{-g} \;\;,
\end{equation}
which contains as dynamical variables $R^{\mu}(\tau,\sigma)$ and
$\xi^{a}(\tau,\sigma)$ a part of which can be removed by gauge fixing.
This action can be thought of as a candidate to qualitatively describe
excitations of slightly curved vortex with winding number 2 due to its
internal multivortex structure.

\section{ Conclusion }

  We have investigated zero modes travelling with a speed of light which are
due to the well known degeneracy of the general planar n-vortex solution in
the Bogomol'nyi limit due to the existence of 2n free parameters. The 2n zero
modes are independent except of a restriction that they all have to be
exclusively left-movers or right-movers. When we identify 2n parameters
with positions of vortices we can say that the general excitation looks like
n vortices which are independently translationally excited. This is a
nontrivial result because there is no restriction as to the distance
between separate vortices.

  What happens when we go outside of the Bogomol'nyi limit ?
For well separated vortices we expect this picture to be preserved. For
distances comparable with mass scales of the theory vortices will attract
or repel depending on whether the strenght of the nonlinearity of the Higgs
field is smaller or greater then in the Bogomol'nyi limit \cite{rebbi,shah}.
When they repel they tend to be well separated but when there are atractive
forces between them they will collapse to one n-vortex configuration.
Small excitations of such a configuration at critical coupling were described
in sections 2.1 and 2.2. The dispersion relation in the Bogomol'nyi limit
for the splitting modes is $k_{i}k^{i}=0$. From a recent work by
P.A.Shah \cite{shah} on extension of the Manton's approximation
a little outside
of the critical coupling one can extract that since for small $\lambda$-s
the nonrelativistic potential is of the harmonic oscilator type the dispersion
relation for small $\lambda$-s will be $k_{i}k^{i}=m^{2}_{\lambda_{k}}$.
Thus we can expect the splitting modes to be preserved but this time with a
massive dispersion relation. It would be interesting to incorporate them
into the newly invented by T.Dobrowolski membrane description
of a vortex with a higher winding number \cite{dobrowolski}. Our examples
of particularly symmetric splitting modes in the end of section 2 will be
modified in this case. Examples a) and b) will be preserved but with
a massive $k_{i}k^{i}=m^{2}$ since they are sinus waves when expressed
in $\lambda$-s. This time they will move up or down the $z$-axis with
a superluminal phase velocity. The twist of c) no longer exists as a coherent
state.

 $Aknowledgement.$ I would like to thank Prof.H.Arod\'z and T.Dobrowolski
for discussions and the staff of the Institute for Theoretical Physics,
Utrecht University, where part of this work was done, for warm hospitality.
I also aknowledge TEMPUS studentship which enabled my stay
in Utrecht and KBN grant 2P302 049 05.


\thebibliography{56}

\bibitem{nielsen} H.B.Nielsen and P.Olesen, Nucl.Phys.B 61 (1973) 45,
\bibitem{vachaspati1} Vachaspati and T.Vachaspati, Phys.Lett.B 238 (1990) 41,
\bibitem{vachaspati2} D.Garfinkle and T.Vachaspati, Phys.Rev.D 42 (1990) 1960,
\bibitem{khariton} V.B.Svetovoy and N.G.Khariton, Yad.Fiz. 48 (1988) 1181,
                                                  Phys.Lett.B 286 (1992) 53,
\bibitem{bogomolny} E.B.Bogomol'nyi, Yad.Fiz. 24 (1976) 449,
\bibitem{weinberg} E.J.Weinberg, Phys.Rev.D 19 (1979) 3008,
\bibitem{taubes} C.H.Taubes, Comm.Math.Phys. 72 (1980) 277, 75 (1980) 207,
\bibitem{ruback} P.J.Ruback, Nucl.Phys.B 296 (1988) 669,
\bibitem{samols} T.M.Samols, Phys.Lett.B 244 (1990) 285,
                             Comm.Math.Phys. 145 (1992) 149,
\bibitem{rubsam} P.J.Ruback and E.P.S.Shellard, Phys.Lett.B 209 (1988) 262,
\bibitem{rebbi} E.Myers, C.Rebbi and R.Strilka, Phys.Rev.D 45 (1992) 1355,
\bibitem{ja} J.Dziarmaga, to appear in Phys.Rev.D,
\bibitem{burzlaf} J.Burzlaff and P.McCarthy, J.Math.Phys. 32 (1991) 3376,
\bibitem{shah} P.A.Shah, Cambridge Univ. preprint DAMTP 94-8,
\bibitem{dobrowolski} T.Dobrowolski, Jagellonian Univ. preprint TPJU-27/93.
\end{document}